\documentclass[nofootinbib]{article}
\usepackage[total={6.5in,9in},top=1in,headsep=0.1in,headheight=1in]{geometry}
\usepackage[utf8]{inputenc}
\usepackage{hyperref}
\usepackage{amssymb}
\usepackage{graphicx}
\usepackage{lipsum}
\usepackage{hyperref}

\usepackage{xspace}

\usepackage{upgreek}
\hypersetup{hidelinks}
\usepackage[hybrid]{markdown}
\usepackage{color}

\newcommand\snowmass{
\begin{center}
  \rule[-0.2in]{\hsize}{0.01in}\\
  \rule{\hsize}{0.01in}\\
  \vskip 0.1in
  Submitted to the Proceedings of the US Community Study\\ 
  on the Future of Particle Physics (Snowmass 2021)\\
 
  \rule{\hsize}{0.01in}\\
  \rule[+0.2in]{\hsize}{0.01in}\\[-2em]
\end{center}
}

\usepackage[firstpage=true]{background}
\backgroundsetup{contents={\parbox{6.5in}{\snowmass}}, scale=1,placement=top,opacity=1,color=black,position={3.25in,1.2in}}

\usepackage{fancyhdr}
\fancypagestyle{plain}{%
  \fancyhf{}%
  \fancyhead[C]{}
  \fancyfoot[C]{\thepage}
}

\fancypagestyle{empty}{%
  \fancyhf{}%
  \fancyhead[C]{{\it Report of the Topical Group on Wave Dark Matter for Snowmass 2021}}
  \fancyfoot[C]{\thepage}
}
\title{Report of the Topical Group on Wave Dark Matter \\ for Snowmass 2021}
\date{}

\usepackage{authblk}

\author[1]{\textbf{Conveners:} Joerg Jaeckel}

\author[2]{Gray Rybka}

\author[3]{Lindley Winslow}

\affil[1]{Institut f{\"u}r Theoretische Physik, Universit{\"a}t Heidelberg, 69120 Heidelberg, Germany}
\affil[2]{Department of Physics, University of Washington, Seattle WA 98195, USA}
\affil[3]{Laboratory for Nuclear Science, Massachusetts Institute of Technology, Cambridge MA 02139, USA}

\newcommand{\DMR}{DMRadio}

\newcommand{\DMRm}{\DMR-m$^3$\xspace}

\newcommand{\DMRP}{\DMR-Pathfinder\xspace}

\newcommand{\micro}{\ensuremath{\mu}}

\begin{document}

\maketitle

\begin{abstract}
There is a strong possibility that the particles making up the dark matter in the Universe have a mass below 1~eV and in many important situations exhibit a wave-like behavior. Amongst the candidates the axion stands out as particularly well motivated but other possibilities such as axion-like particles, light scalars and light vectors, should be seriously investigated with both experiments and theory. Discovery of any of these dark matter particles would be revolutionary. The wave-like nature opens special opportunities to gain precise information on the particle properties a well as astrophysical information on dark matter shortly after a first detection. To achieve these goals requires continued strong support for the next generations of axion experiments to probe significant axion parameter space this decade and to realize the vision of a definitive axion search program in the next 20\,years. This needs to be complemented by strong and flexible support for a broad range of smaller experiments, sensitive to the full variety of wave-like dark matter candidates. These have their own discovery potential but can also be the test bed for future larger scale searches. Strong technological support not only allows for the optimal realization of the current and near future experiments but new technologies such as quantum measurement and control can also provide the next evolutionary jump enabling a broader and deeper sensitivity. Finally, a theory effort ranging from fundamental model building over investigating phenomenological constraints to the conception of new experimental techniques is a cornerstone of the current rapid developments in the search for wave-like dark matter and should be strengthened to have a solid foundation for the future. 
\end{abstract}

\newpage

\tableofcontents 

\newpage
\section{Executive Summary}
The search for dark matter (DM) is one of the great undertakings of particle physics and cosmology. Dark matter is a critical ingredient in shaping the structure of the universe we observe. Yet, we still do not know which new particle it is made from. Candidates for dark matter can be roughly categorized by their mass, with an incredible range from $\sim 10^{-20}$~eV to above the Planck scale $\sim 10^{27}$~eV still being possible. When DM candidates masses fall below 1\,eV, their interactions become wave-like rather than particle-like. The detection techniques are inherently different than traditional particle detectors and they are intrinsically quantum in their nature. Advances in quantum sensing and control along with the related advancements in cryogenics and superconducting magnets have driven an explosion of interests in these candidates. 

The direct detection of any dark matter candidate would clearly be a monumental step forward for both cosmology and particle physics and further strengthen the connection between the two disciplines.
But discovery of a wave-like dark matter particle would bring with it an especially broad range of opportunities. The nature of most experiments searching for wave-like dark matter is such that there is a particularly small gap between what is needed for a discovery and what is needed for precision measurements of properties (e.g. the mass) of the dark matter particle as well as astrophysical information (e.g. the local dark matter velocity distribution). A discovery would start a revolution in particle physics as well as open the field of dark matter astronomy.

Within wave-like dark matter, there is a broad sea of candidates anchored by the highly-motivated QCD axion. An energetic and growing community focused on the search for wave-like dark matter has put forward two goals:
\begin{enumerate}
\item \textbf{Execute a Definitive Search Program for the QCD Axion}  The QCD axion is the theoretically most-studied and strongest motivated WLDM candidate.  Decades of experimental work along with advances in quantum measurement technologies has put us in the unique position:  this decade we can build a suite of experiments that, together, are sensitive to the most plausible theoretical predictions of QCD axion couplings at nominal dark matter densities.  The community intends moving from building technology demonstrators to building machines designed for a discovery.

\item \textbf{Pursue a Theory and R\&D program to elucidate the opportunities in Scalar/Vector Dark Matter}  We are in the process of understanding how WLDM dark matter candidates beyond the QCD axion can work.  There are already experimental techniques that promise to reach previously-unexplored parameters for scalars and vectors.  This Snowmass period we would like to see these experimental techniques refined, and theoretical studies of new WLDM candidates to inform the direction of developing experiments and help them target the most interesting physics.

\end{enumerate}
These goals are in-line with those outlined by the larger dark matter community in the BRN for Dark Matter New Initiatives (DMNI) report. 

The enthusiasm for the QCD axion stems from its key role in solving the Standard Model's strong CP problem (one of the two severe fine tuning problems of the Standard Model) while simultaneously being an excellent dark matter candidate. The QCD axion encompasses two broad classes of models the KSVZ axion which predicts additional quarks and the DFSZ axion which predicts an expanded Higgs sector. Additional  pseudo-scalars called Axion-Like-Particles (ALPs) are a natural consequence of many candidates for fundamental extensions of the Standard Model, in particular string theory. 

Fundamental theories also motivate additional scalar and vector particles which make excellent dark matter candidates. The rich phenomenology of these candidates leads to a variety of detection mechanisms and constraints. Like axions, the wave-like nature of the candidates demands experiments designed around quantum techniques. A discovery of a new scalar or vector would similarly drive a larger high energy physics program to understand the nature of the new physics.

In support of these goals the community has put forward a road map for discovery.
\begin{enumerate}
    \item
    \textbf{Pursue the QCD Axion by Executing the Current Projects:}
    The ADMX G2 effort continues to scan exciting axion dark matter parameter space and the experiments identified by the DMNI process DMRadio-m$^3$ and ADMX-EFR are prepared to start executing their project plans.  These will probe some of the presently most exciting mass ranges with excellent chance of discovery if the axions make up dark matter.
    \item
    \textbf{Pursue Wave-like Dark Matter with a Collection of Small-Scale Experiments} 
    The wave-like nature of these candidates demands the full axion mas range be explored by a collection of significantly different techniques.  These techniques vary in readiness level, but the entire range needs to be comprehensively explored through some combination of them. The DOE DMNI process is the right scale for many of these proposed experiments and was effective at identifying those that were ready to proceed to full projects.  In addition, we should pursue opportunities to harness key US expertise for International projects.

    \item
    \textbf{Support Enabling Technologies and Cross-Disciplinary Collaborations:} Many of the proposed efforts share needs in ultra-sensitive quantum measurement and quantum control, large, high-field magnets, spin ensembles, and sophisticated resonant systems. These technologies overlap strongly with other HEP efforts and synergies should be exploited both within HEP and beyond.
    \item
    \textbf{Support Theory Beyond the QCD Axion:} The QCD axion is an important benchmark model, but not the only motivated one. Theoretical effort should be supported to understand the role of scalars, vectors and ALPs in dark matter cosmology and astrophysics.
\end{enumerate}

The wave-like dark matter experiments enabled by technological advances especially in quantum measurement and control are poised for a great discovery. The US is a leader in this growing field and now is the time to continue the momentum and move this program forward rapidly.

\newpage
\section{Introduction}
 The search for Dark Matter has been a main driver for High Energy Physics (HEP) for several decades. Wave-like dark matter candidates have masses less than 1\,eV. In the local environment with a typical dark matter density ($\rho_{dm}\sim 0.45 {\rm MeV}/{\rm cm}^3$), this corresponds to wavelengths on the order $1\,{\rm km}(0.2\mu{\rm eV}/m)$ and large occupation numbers $\sim 2\times 10^{34}(0.2\mu{\rm eV}/m)^4$. These candidates have been relatively unexplored even though the highly motivated QCD axion is among their number.  The wave-like behavior itself has been the limiting factor because it requires detection mechanisms that are significantly different than traditional High Energy Physics (HEP) or WIMP direct detection experiments. It also means that we have crossed the wave-particle divide and these experiments are intrinsically quantum in their nature. 

Advances in quantum measurement, control, and other enabling technologies including large, high-field magnets, spin ensembles, and sophisticated resonant systems have opened up a new candidates and parameter space to explore. This has inspired a growing community to harness these advancements for HEP and the search for Dark Matter. The community has embraced the Snowmass process and submitted over 86 Letters of Interest and came together to write two community white papers:
\begin{itemize}
    \item \href{https://arxiv.org/abs/2203.14923}{Axion Dark Matter}~\cite{Adams:2022pbo}
    \item \href{https://arxiv.org/abs/2203.14915}{New Horizons: Scalar and Vector Ultralight Dark Matter}~\cite{Antypas:2022asj}
\end{itemize}
They form a complete report on all efforts, from the ongoing DOE Project ADMX-G2 to demonstration-scale experiments to critical R\&D on quantum sensing.

In this Cosmic Frontier topical group (CF2) report we will reiterate (drawing heavily on the above mentioned white papers) the strong motivation for these candidates and build the case for why the goals of this community to
\begin{enumerate}
\item \textbf{Execute a Definitive Search Program for the QCD Axion}
\item \textbf{Pursue a Theory and R\&D program to elucidate the opportunities in Scalar/Vector Dark Matter}
\end{enumerate}
should resonate across high energy physics with a focus on the complementary to the Energy, Rare \& Precision Measurement, and other efforts on the Cosmic Frontier. But there are further strong synergies with Instrumentation Frontier and even with the Accelerator Frontier.

We will review the strong motivation of the QCD
axion and highlights of its role in both particle physics and cosmology in  Sec.~\ref{sec:axion}, and do the same for scalars and vector wavelike dark matter in Sec.~\ref{sec:scalarvector}. 

The current ``big'' projects are briefly discussed in Sec.~\ref{sec:bigprojects}, whereas the role and importance of a diverse range of small projects is stressed in Sec.~\ref{sec:smallproject}. Both big and small projects will benefit from a wide range of enabling technologies as well as a broad and vigorous theory program, Secs.~\ref{sec:enabling} and \ref{sec:theory}. We make our final conclusions in Sec.~\ref{sec:conclusions}.

\section{Definitive Search Program for the QCD Axion}\label{sec:axion}


\begin{figure}[t]
\centering
\includegraphics[trim={0mm 100mm 10mm 75mm},clip, width=0.95\textwidth]{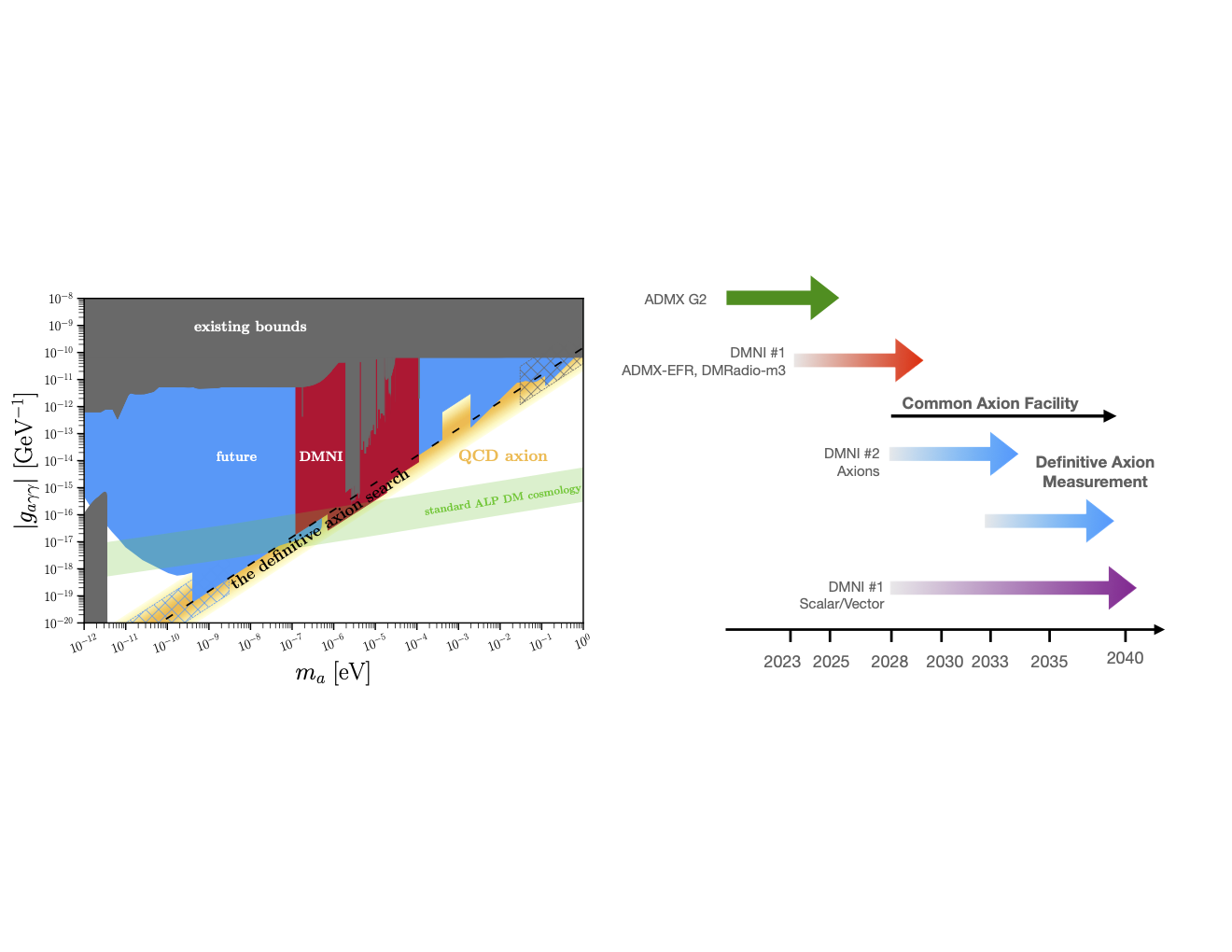}
\caption{(Left) Sensitivities of current and near future axion searches (labelled outlines) compared to the couplings expected for the QCD axion (yellow band) and already excluded areas (grey). Figure by A. Berlin. (Right) Timeline for the experiments and possible common axion facility. Figure by L. Winslow}
\label{fig:axion_summary}
\end{figure}

\subsection{Axions in Particle Physics and Cosmology: Motivation and Role}
Axions take an outstanding role amongst the dark matter candidates as they are motivated by an intrinsic problem of the Standard Model, the strong CP problem, that exists independently from the need for DM. The strong CP is one of two severe fine-tuning problems of the Standard Model (the other being the smallness of the Higgs vacuum expectation value compared to the Planck scale). To be consistent with constraints on the neutron electric dipole moment~\cite{Abel:2020pzs} the CP violating dipole moment requires a tuning of 1 in $\sim 10^{10}$ of the so-called $\theta$ angle.
In contrast to other fine-tunings in the SM this problem is unique as it cannot be addressed by anthropic reasoning. We could almost certainly exist in a form very similar to our present one even for a $\theta\sim 0.1$~\cite{Lee:2020tmi}.

This tuning problem is solved by the famous Peccei-Quinn mechanism~\cite{Peccei:1977hh,Peccei:1977ur} that is an important early example of solving a fine-tuning problem by dynamical relaxation. The $\theta$-parameter is turned into a dynamical field that then over the cosmological evolution relaxes to its potential minimum located at the CP conserving value that ensures a vanishing electric dipole moment of the neutron.
The crucial observable consequence of $\theta$ being dynamical is that its excitations correspond to particles, i.e.axions~\cite{Weinberg:1977ma,Wilczek:1977pj}.
An important aspect of the PQ solution is that it still fully addresses the fine-tuning problem. This is in contrast to proposed solutions of the electroweak hierarchy problem, such as, e.g. supersymmetry, where experimental constraints have led to the reemergence of a sizable amount of tuning (cf., e.g.~\cite{Jaeckel:2012aq,Feng:2013pwa,vanBeekveld:2019tqp}).

In the range $f_{a}\sim (10^{10}-10^{13})\,{\rm GeV}$ the misalignment mechanism~\cite{Preskill:1982cy,Abbott:1982af,Dine:1982ah,Turner:1983he,Turner:1985si} provides a good way for axions to be the cold dark matter in the Universe.  This mechanism is an automatic consequence of the dynamical relaxation of the $\theta$-angle. If the initial $\theta$-angle is non-vanishing, this corresponds to a non-vanishing energy density. This  energy density is diluted by the cosmic expansion, reducing the $\theta$-angle to its potential minimum at 0. It is straightforward to check that this energy density behaves exactly as one would require for cold dark matter. Therefore the axion is a natural dark matter candidate in the sense that unless tuning of the initial value, or a strongly non-standard cosmology is applied it inevitably yields an amount of dark matter roughly of the order of he observed size\footnote{Additional contributions of a similar size may arise from topological defects in the case that the final Peccei Quinn symmetry breaking happens after inflation~\cite{Harari:1987ht,Davis:1989nj,Dabholkar:1989ju,Hagmann:1990mj,Battye:1993jv,Yamaguchi:1998gx,Chang:1998tb,Hagmann:2000ja,Sikivie:2006ni,Hiramatsu:2010yn,Hiramatsu:2010yu,Hiramatsu:2012gg,Hiramatsu:2012sc,Kawasaki:2014sqa,Gorghetto:2018myk,Martins:2018dqg,Hindmarsh:2019csc,Gorghetto:2020qws,Dine:2020pds,Hindmarsh:2021vih,Buschmann:2021sdq}. These affect the argument only on a quantitative level.}. Moreover, this dark matter is automatically (very) cold, as required by structure formation.


\subsection{Axion DM discovery: What can we learn? - Particle Physics}

The discovery of a new fundamental particle is in any case an enormous step forward for physics. 
For the axion this would be especially true. The axion is automatically connected to an energy scale $f_a\gtrsim 10^{7}\,{\rm GeV}$. Thereby its pure existence gives a glimpse at fundamental physics at scales more than 1000 times larger than those directly accessible at colliders such as the LHC.  
But even beyond that, its nature as wave-like dark matter would allow for high precision measurements soon after discovery that will give important information both for fundamental particle physics (this subsection) as well as astrophysics and cosmology (next subsection).

Let us briefly consider a few of the measurements that would be possible relatively soon after an initial discovery, the insights that can be gained from them, as well as their connection to other areas of particle physics.

The currently most sensitive axion dark matter search experiments use the wave-like nature and in particular the intrinsically long coherence length of the axion wave to achieve a resonance between the oscillations of the axions with a frequency $m_{a}$ and suitable resonator, e.g. a cavity. In this way the detection of axions is enhanced by a factor of the quality factor of the resonator up to $Q\sim 1/v^2\sim 10^6$. Of course, this enhancement only applies within the resonance width $\sim 1/Q$, i.e. a mass within a window $\Delta m_{a}\sim m_{a}/Q$ around the known eigenfrequency of the resonator, $\omega_{\rm res}=m_{a}$. Therefore, essentially the detection itself automatically yields a measurement of the axion mass with an up to $10^{-6}$ precision.

For a standard QCD axion, measuring the mass immediately implies a measurement of the corresponding axion decay constant, i.e. $f_a$, via the relation~\cite{Weinberg:1977ma,Wilczek:1977pj,GrillidiCortona:2015jxo}\footnote{The precision in the expression being limited by the precision of the underlying QCD/chiral perturbation calculation. Consequently, improvements in the relevant QCD input could also make this more precise, giving also a first example of the synergy between the two areas.},
\begin{equation}
    f_{a}\approx 10^{10}\,{\rm GeV}\left(\frac{0.6{\rm meV}}{m_{a}}\right).
\end{equation}
We would therefore get direct evidence for a new large energy scale in particle physics.
This could open connections to a wide range of fundamental structures.
For example, in string theories the axion scale $f_a$ is usually directly linked to the string scale~\cite{Svrcek:2006yi,Conlon:2006ur, Arvanitaki:2009fg, Cicoli:2012sz, Marsh:2015xka, Acharya:2015zfk, Visinelli:2018utg, Broeckel:2021dpz, Demirtas:2021gsq}, but its size is also suggestive of a scale suitable for the see-saw scale in neutrino models as made explicit, e.g. in the SMASH model~\cite{Ballesteros:2019tvf}. Measuring the properties and couplings of the axion will give information on which option is realized in nature, with significant implications for theoretical model building.

A next step would be to actually demonstrate that the discovered dark matter particle, seen, e.g., in an experiment targeting the axion photon coupling (cf., e.g., the experiments discussed in  Sec.~\ref{sec:bigprojects}), is indeed a QCD axion.
To achieve this requires establishing the defining gluon coupling. This could be done by an experiment  directly targeting this coupling, e.g. by searching for an oscillating electric dipole moment of nuclei as done  by CASPEr~\cite{Graham:2013gfa,Budker:2013hfa}. 
Such a measurement would  be strongly facilitated by already knowing the precise mass of the axion, as this eliminates the need for scanning, speeding up the experiment by a factor of up to $Q\sim 10^6$. Or, more importantly in this case, allowing for a significant increase in the measurement  time at  the given frequency, allowing an experiment that would otherwise not be sensitive to a QCD axion to nevertheless make a  detection.\footnote{For the same reason it is likely that several experiments targeting the discovery coupling could confirm the discovery relatively shortly after an initial discovery.}

The photon coupling is usually determined by a the combination of a model dependent electromagnetic anomaly plus a model independent contribution from pion mixing~\cite{Bardeen:1977bd,Kaplan:1985dv,Srednicki:1985xd}. A sufficiently precise determination may allow to infer if the axion in question is a (sufficiently simple) DFSZ~\cite{Zhitnitsky:1980tq,Dine:1981rt} or KSVZ~\cite{Kim:1979if,Shifman:1979if} type axion. 
Even assuming that the axion constitutes all of the dark matter, the local axion density is at present not sufficiently well known to make this distinction. However, this may be different if also a detection in a helioscope~\cite{Sikivie:1983ip} such as IAXO~\cite{Irastorza:2011gs,Vogel:2013bta} is achieved. Alternatively such a distinction can be made if either IAXO or a direct detection experiment such as QUAX~\cite{Irastorza:2011gs,Vogel:2013bta} measure a sizable coupling to electrons, which is expected only in DSFZ type models. This would have crucial implications for collider searches, as DFSZ models feature two Higgs doublets that typically should not be too heavy. They would therefore be within reach of collider searches, possibly even of the LHC. 
\subsection{Axion DM discovery: What can we learn? - Cosmology and Astrophysics}

A crucial question for any dark matter candidate is whether it contributes all or only a part of the dark matter.  For an axion haloscope, the signal is proportional to the product $\sim g^{2}_{a\gamma\gamma}\rho_{a}$ and so the coupling and local density are degenerate with only a single detection.  Fortunately are at least two options to experimentally disentangle them. 
 If a haloscope detection can be combined with a search for a candidate in an experiment that is independent of dark matter, e.g. the helioscope~\cite{Sikivie:1983ip}, IAXO~\cite{Irastorza:2011gs,Vogel:2013bta}, whose signal strength is $\sim g^{4}_{a\gamma\gamma}$ this can be used to infer the local axion dark matter density $\rho_a$ which can then be compared with the expected dark matter density $\rho_{\rm CDM}\sim (300-450)\,{\rm MeV}/{\rm cm}^3$.
If we also have a detection via the gluon coupling/electric dipole moment, we can also use that in this case the size of the coupling is uniquely related to the mass. Knowing the mass we can therefore directly infer the local dark matter density from the strength of the signal amplitude $\sim g_{agg}\sqrt{\rho_{a}}$.
The precision with which this comparison can be achieved in both cases also provides a nice synergy to astrophysical observation and structure formation simulations that give predictions for the local dark matter density. Moreover, precise predictions for a relation between the axion gluon coupling and the resulting oscillating electric dipole moment require theory input from QCD and nuclear physics, giving another important synergy area.

Important astrophysical/cosmological information, in particular on the dark matter distribution and therefore on  small scale structure formation will be available already shortly after an initial discovery.
Resonant experiments such as ADMX, ABRA and others (see Secs.~\ref{sec:bigprojects} and \ref{sec:smallproject}) measure an amplitude of the signal and are able to perform a high precision spectral analysis. Due to the wave-like nature of axion dark matter this spectrum directly corresponds to the energy spectrum of the axions converted in the detector. In that way we can obtain a precise measurement of the kinetic energy, i.e. velocity squared, distribution of axion in our location of the Galaxy. A $\Delta v^2\sim 10^{-9}$ resolution of the velocity squared distribution compared to a typical $v^2\sim 10^{-6}$ is feasible in most discovery experiments without any modification and within a very short timescale (cf.~\cite{Duffy:2005ab} for a high resolution measurement done by ADMX within a data taking campaign).
Observing daily and annual modulations of the resulting spectrum can even give first information on the full vectorial distribution~\cite{PhysRevD.42.3572}. This may be further enhanced by exploiting the (small) intrinsic directionality of several of the detection schemes (e.g., MADMAX~\cite{Millar:2017eoc}, ADMX~\cite{Irastorza:2012jq}).

Additional astrophysical information may be gleaned if IAXO is also able to detect axions\footnote{For this axions do not need to be dark matter.}. In this case the axion may, e.g. also serve as a probe of the solar composition (e.g. metalicity, elemental abundances)~\cite{Jaeckel:2019xpa}.

These examples demonstrate that discovery of wave-like axion dark matter would open a new window of ``dark astronomy''.

\subsection{Beyond QCD Axions: Axion-like particles - Motivation and Discovery}

While the axion is the most prominent wave-like dark matter candidate it is by far not the only one.
An important generalization are so-called axion-like particles, loosely defined as particles sharing crucial features of the axion (e.g. (pseudo-)scalar, low mass, weak couplings) but not solving the strong CP problem.
In field theory they can be imagined as pseudo-Nambu Goldstone bosons arising from the spontaneous breaking of approximate global symmetries.
In string theory they can originate as the imaginary part of moduli fields~\cite{Conlon:2006tq,Conlon:2006ur,Arvanitaki:2009fg,Cicoli:2012sz,Marsh:2015xka,Acharya:2015zfk,Visinelli:2018utg,Broeckel:2021dpz,Demirtas:2021gsq,Mehta:2020kwu,Mehta:2021pwf}. Both are rather generic features of extensions of the Standard Model and therefore we would them to be quite common in many extensions of the Standard Model. Indeed, axion-like particles could also arise in models for the flavor structure of the Standard Model (they are then often also dubbed familons)~\cite{Wilczek:1982rv} and in string theory there has been plenty of discussion on the existence of axion-like particles including the possibility to have a sizable number of them a so-called string axiverse~\cite{Arvanitaki:2009fg}.
Similar to axions they can be wave-like dark matter via the misalignment mechanism~\cite{Preskill:1982cy,Abbott:1982af,Dine:1982ah,Turner:1983he,Turner:1985si,Arias:2012az}.

The mass of axion-like particles is not determined by QCD effects as is the case for QCD axions. Therefore, a much wider range is possible. 
Moreover, a possible strong temperature dependence~\cite{Arias:2012az}, an enlarged field range~\cite{Jaeckel:2016qjp} or coupling enhancing effects~\cite{DiLuzio:2016sbl, DiLuzio:2017pfr, DiLuzio:2020wdo,Plakkot:2021xyx,Sokolov:2021ydn} may allow couplings to be much stronger than that expected for the simplest QCD axions. This gives true discovery potential also  to experiments that do not (yet) reach the QCD axion  band.

As is the case for the QCD coupling to the Standard Model proceeds via higher dimensional operators (typically derivative couplings). Therefore, as in the case of the QCD axion a discovery would indicate a clear scale for new physics. Any such discovery would therefore essentially give rise to a no-loose theorem for experiments that explore this new scale with a broad reach.
Axion-like particles are often less constrained than QCD axions. In particular the couplings could often still be stronger than those  of  QCD axions, indicating a lower physics scale that could be within reach of other experiments. Moreover, a stronger coupling could also put them into the reach of other experiments, e.g. IAXO, that are independent of dark matter. 

Most of the precise post-discovery measurements that could be performed with a dark matter axion (see previous subsections) are likely to be also possible with an axion-like particle. Therefore, an axion-like particles would open a similar level of access to a wealth of information both on the underlying fundamental particle physics model as well as new insights into astrophysics and cosmology.


\section{Wave-like DM scalars and vectors: A New Horizon}\label{sec:scalarvector}

\begin{figure}[h!]
\centering
\includegraphics[trim={0mm 0mm 0mm 0mm},clip, width=0.85\textwidth]{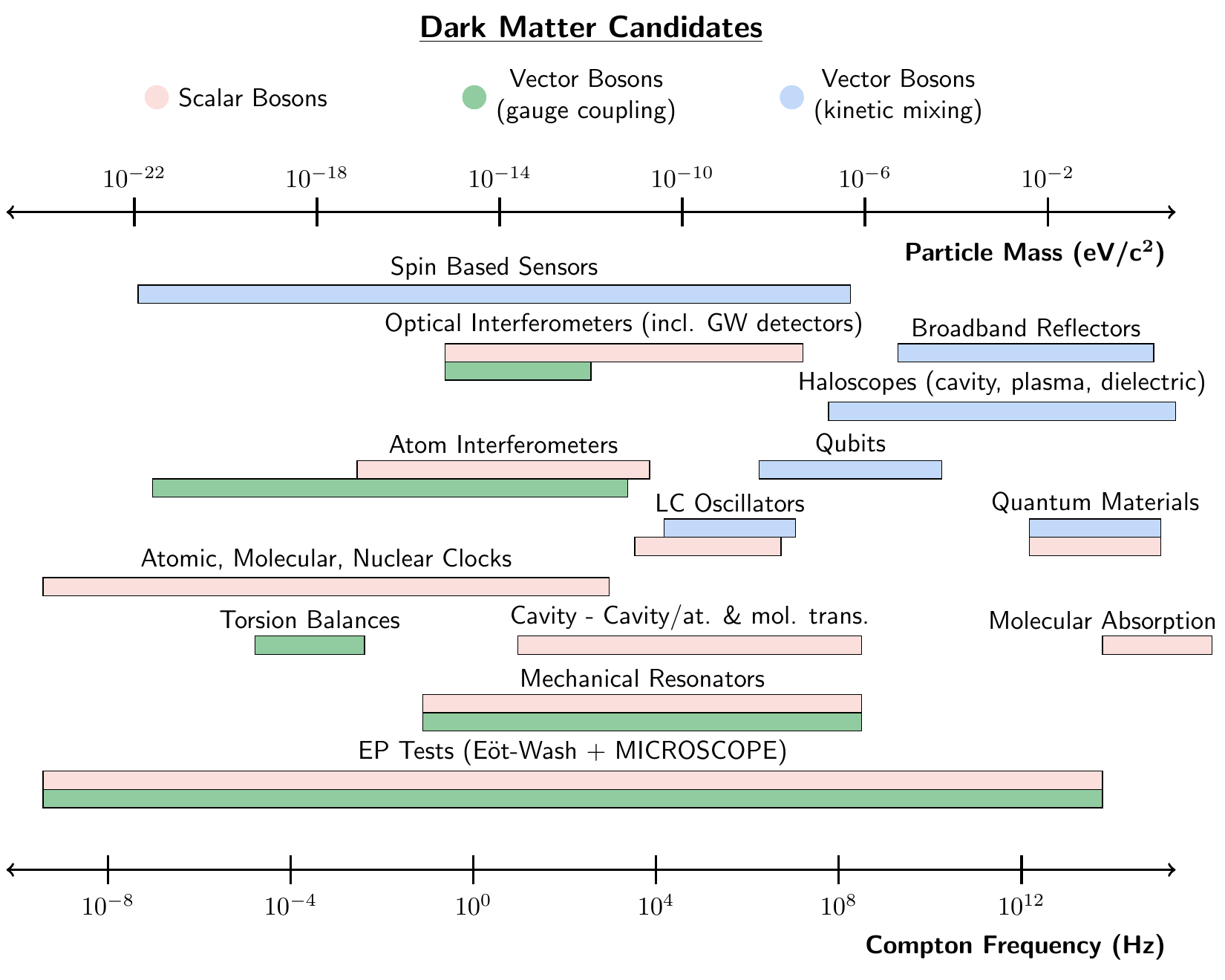}
\caption{Overview of experimental techniques and the mass ranges they target, for scalars (pink) and vector (green and blue). Figure taken and caption adapted from~\cite{Antypas:2022asj}. }
\label{fig:sv_summary}
\end{figure}

While axions (and axion-like particles) are an important benchmark scenario for wave-like dark matter they are far from the only well-motivated candidates.
As wave-like dark matter must be bosonic to avoid the Pauli principle we in general have the possibility to have very light scalars or vectors.
Indeed they are amongst the simplest possible extensions of the Standard Model, by a single scalar and by a simple U(1) gauge factor in the case of vectors.

\subsection{Scalars and Vectors in Particle Physics - Motivation}
A central motivation for scalars and vectors is that they are the only bosonic particles that can interact with the Standard Model via renormalizable ``portals''. For scalars ($\phi$) via the ``Higgs-portal''~\cite{Binoth:1996au,Schabinger:2005ei,Patt:2006fw,Ahlers:2008qc,Batell:2009yf} $(\kappa\phi+\lambda\phi^2)|H|^2$ and in the vector ($X_\mu$) case via ``kinetic mixing''~\cite{Okun:1982xi,Holdom:1985ag,Foot:1991kb} $\chi B_{\mu\nu}X^{\mu\nu}$ where $B_{\mu\nu}$ is the hypercharge field strength (at low energy it's essentially the electromagnetic field strength) and $X_{\mu\nu}$ that of the new vector field.
The  portal interactions are special in that they are not suppressed by some (very) high energy scale. As such they can be generated at a very  high energy scale while still leaving observable traces in low energy experiments.
However, due to the exceptional sensitivity of many low-energy experiments searching for scalars and vectors, also additional interactions (that are suppressed by a high energy scale) can be accessed.

Scalars and vectors are also abundant in extensions of the Standard Model based on string theory. Indeed, string theory usually features plenty of scalars in the form of moduli (the real part), and brane constructions of the  Standard Model generically contain extra U(1) gauge factors giving rise to vector bosons (cf., e.g.,~\cite{Conlon:2006tq,Conlon:2006ur,Arvanitaki:2009fg,Cicoli:2012sz,Marsh:2015xka,Acharya:2015zfk,Visinelli:2018utg,Broeckel:2021dpz,Demirtas:2021gsq,Mehta:2020kwu,Mehta:2021pwf,Hebecker:2022fcx,Dienes:1996zr,Lukas:1999nh,Abel:2003ue,Blumenhagen:2005ga,Abel:2006qt,Abel:2008ai,Goodsell:2009pi,Goodsell:2009xc,Goodsell:2010ie,Heckman:2010fh,Bullimore:2010aj,Cicoli:2011yh,Goodsell:2011wn}).

Scalars also feature in models to address the hierarchy problem (e.g. the relaxion~\cite{Graham:2015cka}) (synergy with theory), and may also play a role in dark energy (e.g. quintessence~\cite{Wetterich:1987fm,Ratra:1987rm}) giving a synergy with CF4-6.

It is noteworthy that very light scalars and vectors can often be tested to an amazing precision, often at a level that is at or even below gravitational  strength (see~\cite{Antypas:2022asj} for a wide variety of examples). This gives access to truly fundamental effects at the highest imaginable energy scales.

\subsection{Scalar and Vector DM - Cosmology}
For both types of particles their role as wave-like dark matter is supported by the existence of suitable production mechanisms that can generate sufficient amounts to explain the full observed dark matter abundance. The mechanisms include the misalignment mechanism~\cite{Preskill:1982cy,Abbott:1982af,Dine:1982ah,Turner:1983he,Turner:1985si,Nelson:2011sf,Arias:2012az} already discussed in the case of axions, but there are also mechanisms that are based on resonant/tachyonic decays~\cite{Dror:2018pdh,Agrawal:2018vin, Co:2018lka, Co:2021rhi,Bastero-Gil:2018uel}, inflationary and gravitational production~\cite{Graham:2015rva} as well as the decay of topological configurations such as strings~\cite{Long:2019lwl}.
Notably, the production mechanism, clumpy vs non clumpy or for vectors the type of polarization can also impact the required search strategies~\cite{Arias:2012az,Graham:2015rva,Alonso-Alvarez:2018tus,Alonso-Alvarez:2019ixv,Caputo:2021eaa}.

\subsection{Scalar and Vector DM - Detection and Discovery}
From the experimental side the search for scalars and vectors is driven from two directions. One is that experiments searching for axions are often also directly sensitive to vectors\footnote{There is also  some sensitivity to certain types of scalars and more might be achievable with adapted configurations~\cite{Flambaum:2022zuq}.}  (indeed the search is often simpler, e.g. by not requiring a superstrong magnetic field to be present).
A second equally important driver are a wide diversity of ultra-precise measurements using an amazing range of experimental techniques and technologies ranging from atomic physics measurements, atomic clocks, ultrasensitive mechanical sensing, optical and atom interferometry to observations of the cosmic microwave background and black holes (e.g. via gravitational waves). All these often employ and further develop cutting edge technologies such as, in particular also quantum technologies.

As in the case of axions, opportunities go far beyond the discovery of a new particle and dark matter. Indeed in most cases the post-discovery measurements discussed above for the case of axions are based on the wave-like nature and can also be done in the case of scalars and vectors. Therefore, in this case, too, unprecedented information on particle physics as well as cosmology would be within reach.

\section{QCD Axion Coverage}\label{sec:qcdcoverage}

One of our goals for the wavelike dark matter community is to ``execute a definitive search for the QCD axion".  We will define a `definitive search' as sensitivity to QCD axions with DFSZ coupling if they make up the majority of our local dark matter density.  We believe this is achievable, but it should be noted that it is not exhaustive; as describe in Sec. \ref{sec:theory} it is conceivable that axions have smaller couplings or make up a minority fraction of the local dark matter density.  In the event of no discovery, such considerations can be addressed in the next Snowmass period.

\begin{figure}[h!]
\centering
\includegraphics[trim={0mm 0mm 0mm 0mm},clip, width=0.85\textwidth]{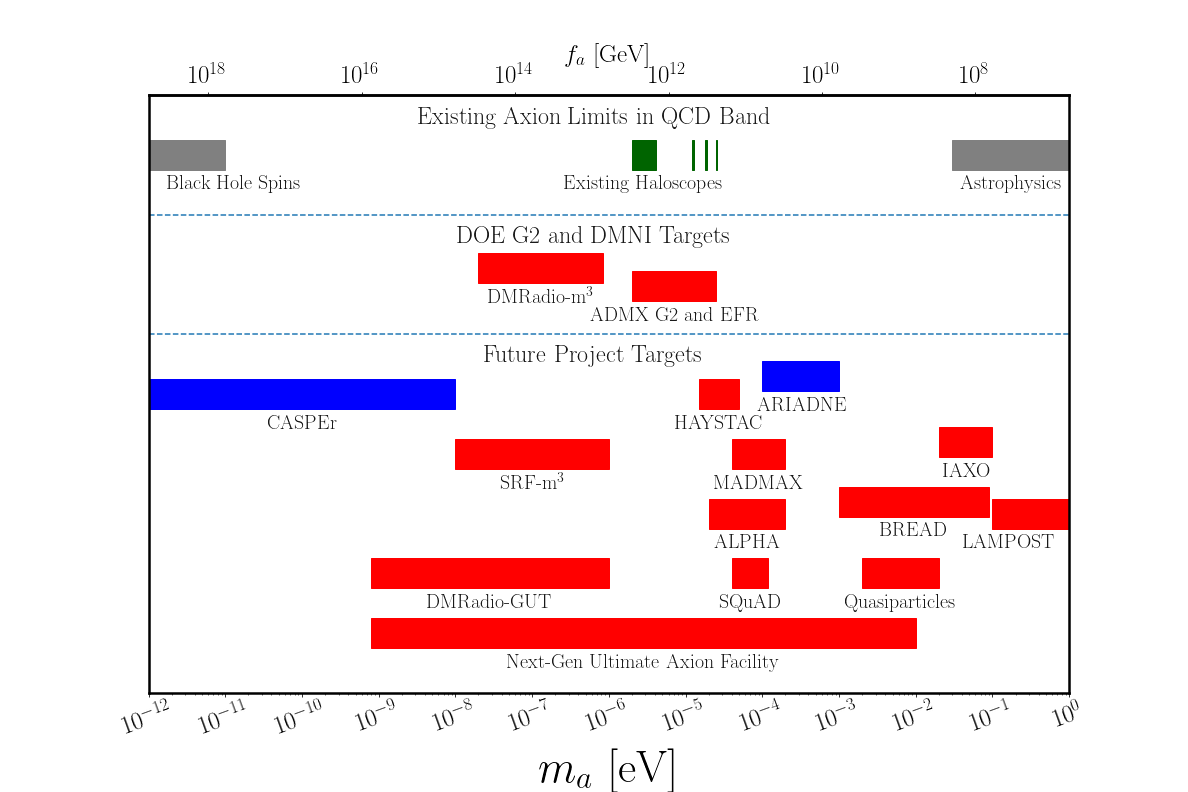}
\caption{Axion mass ranges explored by different experiments. Running experiments are shown in green. The proposals based on the axion-photon coupling are shown in red, those targeting other couplings in blue. Figure taken and caption adapted from~\cite{Adams:2022pbo}. }
\label{fig:current_projects}
\end{figure}

Achieving this goal is not tenable by any one experiment.  Different energy scales require different technologies.  The applicability of current ant future projects over the QCD axion mass range is shown in Fig.~\ref{fig:current_projects}.

Here we have divided coverage into existing limits, current ongoing experiments, and future projects.  The relevant QCD axion masses are bounded from above by  stellar astrophysics, and from below by the non-observation of axion superradiance effects on black hole spins.  A beachhead into the QCD axion couplings in the $\mu$eV has already been established by the ADMX G-2, HAYSTAC, and CAPP cavity haloscopes, demonstrating their effectiveness.  Experiments currently funded will expand this explored region as described in Sec.~\ref{sec:bigprojects}.  A complete exploration of the QCD axion mass range will require supporting R\&D to establish a suite of experiments in the near future as described in Sec.~\ref{sec:smallproject}.


\section{The Current Projects}\label{sec:bigprojects}

\subsection{ADMX-G2}

The ADMX program has led the axion community for many decades. The current stage of the program ADMX-G2 has been running since 2015. It is the first experiment to reach the sensitivity to search for QCD axions  at typical DFSZ coupling making up the bulk of the local dark matter density~\cite{ADMX:2018gho}. The experiment is built around a large (8.5 Tesla) high inductance (540 Henry) NbTi Superconducting Solenoid magnet with a 60\,cm inner diameter and a height of 112\,cm. The induced axion signal the couples to a cavity and is readout with quantum amplifiers. It has performed three runs (\cite{ADMX:2018gho},~\cite{PhysRevLett.124.101303},~\cite{ADMX:2021nhd}). Each run used slightly different cavity geometries to provide sensitivity to increasing frequencies. The quantum amplifier technology has been key. Run 1A used a Microstrip SQUID Amplifier but subsequent runs used Josephson Parametric Amplifiers (JPAs) for amplification. The final run will combine 4 frequency locked cavities coherently to take advantage of the axions coherent signal to make up for the volume lost by operating a single cavity. This is scheduled to complete its scan as the DMNI project ADMX-EFR completes its construction and commissioning, as indicated in Figure~\ref{fig:axion_summary}. ADMX-G2 will have succeeded in search approximately two octaves of parameter space from  2$\mu$eV to 8$\mu$eV. Figure~\ref{fig:current_projects} shows in green the current the ADMX-G2 accomplishments and the current status of demonstrator-scale experiments HAYSTAC and CAPP which are now probing QCD axions at near-KSVZ sensitivity. 

\subsection{DMNI Project: ADMX-EFR}

The ADMX collaboration aims to push to the higher frequency range of 2{\textendash}4 GHz with the Extended Frequency Range (EFR) experiment. The goal is to achieve DFSZ sensitivity across this frequency range, which corresponds to an axion-mass range of 8.3{\textendash}16.5 $\mu\mathrm{eV}$. ADMX plans to overcome the sensitivity loss due to the decreasing volume of the cavity operating an array of eighteen cavities. Further sensitivity improvements come from increased magnetic field strength from an existing 9.4\,T MRI magnet, increased quality factor by coating cavities with superconducting films or low-loss dielectrics and finally reduced amplifier noise enabled by squeezed state amplifiers or single photon counting techniques that evade the Standard Quantum Limit. The experiment is ready to make great inroads into critical parameter space as soon as the DMNI projects are given the green light. The ideal funding scenario would have the run start in 2024 completing 2027.

\subsection{DMNI Project: DMRadio-m$^3$}

The \DMRm experiment aims to search for the QCD axion over the mass range $20\,\mathrm{neV}\lesssim m_a\lesssim 800\,\mathrm{neV}$, achieving DFSZ sensitivity over the range $120\,\mathrm{neV}\lesssim m_a \lesssim 800\,\mathrm{neV}$ with a 5 year total scan time. This mass range begins to probe the top of the GUT-scale QCD axion masses, in addition to testing a wide swath of interesting ALP parameter space. In this mass and frequency range (30\,MHz to 200\,MHz), the axion signal starts behaving like a current, so the experiment can started to be modelled as lumped element circuit components, hence these experiments are referred as the lumped element experiments in comparison to the cavity experiments like ADMX. Like ADMX-EFR, \DMRm is ready to search a wide complementary parameter space. In the ideal funding scenario, construction starts in 2024 with the physics data taking from 2026-2031.

\subsection{Current Demonstrators and Future Projects}

As will be described in Section~\ref{sec:smallproject}, the nature of wave dark matter detection requires an evolution of techniques as a function of frequency and candidate. For axion physics we outline in Figure~\ref{fig:axion_summary} many of the efforts that are demonstrating many of the key technologies and techniques outlined in Section~\ref{sec:enabling} that will be critical for the definitive search for the axion. We highlight in blue the techniques that are pursuing the measurements of the alternative couplings which will be particularly critical in the event of a detection. 


\section{Importance of Small Projects} \label{sec:smallproject}

The wave-like nature of candidates in the sub-1\,eV mass range leads to detection mechanisms that are inherently quantum and require detection schemes that are very different than standard particle detectors. The wave-like nature also demands experiments optimized for the frequency range of interest. Much intuition can be gleamed from our understanding of the detection of electromagnetic radiation, the techniques for measuring radio waves are very different than visible light. For this reason the search for wave-like dark matter demands a suite of experiments not one monolithic detector. This suite of experiments requires a strong program of experiments at different scales from R\&D, see Section~\ref{sec:enabling} to demonstrators to Small Projects. 
 
 \subsection{Demonstration-Scale Experiments}

The current suite of experiments has a healthy mix of experiments. Having a range of demonstration-scale experiment has been key and in light of the huge mass/coupling range that needs to be tackled it will be very important to also have strong support for new initiatives at this level. In axions, demonstration-scale Experiments on the order of $\sim$\$100k to $\sim$\$1M  such as ABRACADABRA, \DMRP, ADMX-SLIC, ADMX-Orpheus, ADMX-Sidecar and HAYSTAC provided key results that proved readiness to move to full projects. These were funded as a combination of Foundation money and grants to individual institutions. Support of demonstrators continues to be key part of the pipeline, establishing new approaches and technologies. We highlight experiments like CASPEr to measure the nuclear couplings and  HAYSTAC which continue to innovate in quantum sensor readout. A variety experiments in scalar/vector that will move from table-top R\&D to demonstrators and we should be poised to embrace these opportunities.

\subsection{DMNI Process and Small Projects}
In 2018, the dark matter community came together to write a Basic Research Needs Report (BRN) for Dark Matter New Initiattives (DMNI). This outlined general goals for the field. This was followed by a DOE funding call to provide funds for the development of project execution plans for full proposals. The selected experiments,\DMRm and ADMX-EFR, were described in Section~\ref{sec:bigprojects}. 

The current DMNI experiments are ready to progress to construction and will commence operations in the next five years. To keep an active pipeline, a new DMNI call for proposals to prepare projects should come concurrently with the current experiment proceeding to construction.

\subsection{Axion Facility}

For axion experiments, much of the infrastructure to host experiments could be common. This includes electromagnetic shielding, cryogenic systems and low noise warm electronics with well-characterized grounding schemes. This infrastructure may be appropriate for scalar/vector searches depending on the details of the detection technique. The magnet is the cost-driver for many of the experiments. It may also be possible for the experiments in the $>$~GHz regime to share one smaller bore high-field magnet while in the $<$~MHz regime share a large magnet with slightly more modest fields. The community is interested in coming together to produce a conceptual design for 1-2 such facilities in support of future small projects.

Indeed in Germany and South Korea efforts for a consolidation are already underway with the plan to house three  experiments at DESY (ALPS-II, IAXO and MADMAX) as well as the construction of a whole suite of dark matter experiments at CAPP. In both cases experiments strongly benefit from existing and expanding expertise and infrastructure in particular with regards to magnets and cryogenics. 

Having such a facility at hand could significantly speed up (and reduce  costs) further development as well as construction of experiments and help to achieve a leading role for the US program.

\subsection{Work Force Development}

Demonstrator-Scale experiments and Small Projects are incredibly good for training the next generation of experimentalists.
In such experiments, students and postdocs get experience in all aspects of the experiment. The training of instrumentalists has been highlighted as a strategic need for the field. Wave dark matter experiments are particularly valuable training grounds since the enabling technologies including quantum sensing and control, as described in Section~\ref{sec:enabling},  have been identified as national priorities. For these reason, wave dark matter experiments are particularly critical as drivers for work force development.

\section{Enabling Technologies} \label{sec:enabling}

The small projects described in Section~\ref{sec:smallproject} are enabled by a strong R\&D program. In this section, we highlight the technologies that are at the heart of searches for wave-like dark matter. This is not exhaustive as many new techniques are constantly emerging beyond the horizon of axions, ALPs, scalars and vectors summarized in Figures~\ref{fig:axion_summary},~\ref{fig:sv_summary},~\ref{fig:current_projects}. We emphasize that these are very much interdisciplinary activities involving cross frontier topics such as magnet and cavity development, but also groups that may classically be atomic physicists who bring key experience in quantum control and measurement.  

\subsection{Quantum Measurement and Control}

By definition the search for wave-like dark matter sits at the transition from particles behaving like particles and them behaving like waves, therefore the detection techniques are inherently quantum and quantum measurement and control is a key enabling technology.  This directly aligns with recent national initiatives to increase investment in quantum technology and build the quantum workforce of the future.

\subsubsection{Moving Beyond the Standard Quantum Limit}

\begin{figure}[t]
\centering
\includegraphics[trim={25mm 50mm 25mm 50mm},clip, width=0.85\textwidth]{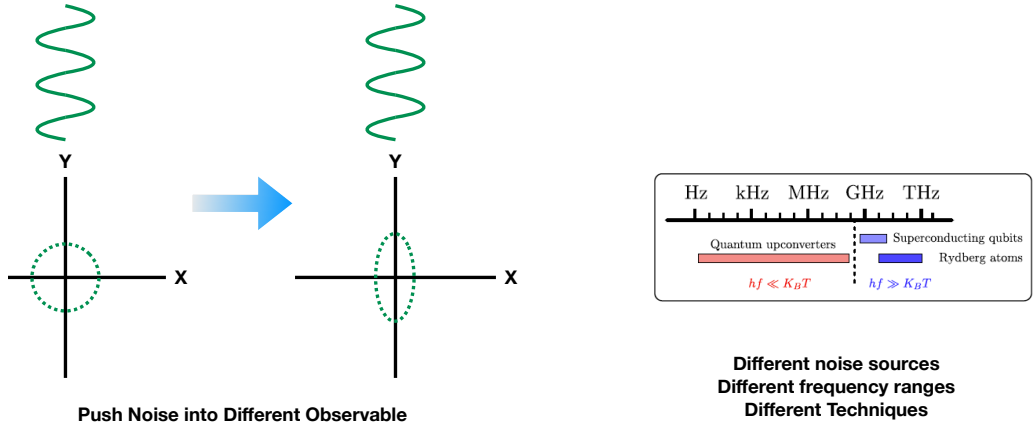}
\caption{Principles of beyond the standard quantum limit amplification. (Left) The goal is to push the quantum noise out of the experimental observable. (Right) The source of quantum noise changes as a function of frequency and therefore the techniques change. Figure by L. Winslow with A. Chou and K. Irwin}
\label{fig:quantumtechniqes}
\end{figure}

For the detection of wave-like dark matter, the experiment ultimately couples to an electromagnetic sensor. As the temperature of the experiment is reduced below a few hundred milli-kelvin, the sensitivity of the experiment is dominated by quantum noise rather than thermal noise. To improve sensitivity, the sensor needs to push beyond this standard quantum limit. This is achieved by pushing the noise out of the signal being measured as indicated in Figure~\ref{fig:quantumtechniqes}. The details of how to achieve this are frequency dependent. Below about 1\,GHz, the noise is quantum back action and the corresponding techniques is back action evasion using devices such at quantum upconverters. Above 1\,GHz the statistics of the photons dominates and techniques such as quantum squeezing can be used to evade the standard quantum limit. Quantum squeezing was successfully demonstrated by the HAYSTAC collaboration to achieve a quantum advantage gain of about 2.5. Single photon counting promises an even larger quantum advantage \cite{lamoreaux2013analysis} and a superconducting qubit-based single photon detector was recently used in a dark photon search using a fixed frequency cavity \cite{Dixit:2020ymh}. These devices must be coupled with high Q resonators or cavities ($Q > 5\times 10^{5}$).

\subsubsection{Engineering Spin Ensembles}


Quantum engineering of spin ensembles are used in experiments searching for all types of wave-like dark matter. These include experiments such as CASPEr which can search for EDM and gradient interactions induced by both axions and scalar dark matter candidates~\cite{Aybas:2021nvn}.  The goal is to observe spin ensemble dynamics at the level of spin projection noise and then use spin squeezing and more generally spin ensemble correlations to further increase sensitivity. The spin ensemble is ultimately coupled to an electromagnetic sensor so the techniques developed above will find application here. The other key R\&D activity is engineering to optimize the materials that host these spin ensembles to increase the sensitivity to the EDM and gradient interactions of ultra-light dark matter~\cite{Budker2014}.

\subsubsection{Atomic Clocks}

There are many novel and interesting techniques in development for the search for scalar and vector dark matter, see Figure~\ref{fig:sv_summary}. We highlight here atomic clocks as due to its broad application. In the last $\sim15$\,years, optical atomic clocks have improved by more than three orders of magnitude in precision,reaching a fractional frequency precision below 10$^{-18}$ \cite{2019Alclock}. The interaction of scalar dark matter can lead to the oscillation of fundamental constants such as the fine-structure constant $\alpha$ or proton-to-electron mass ratio. If these vary in space or time then atomic, molecular or nuclear will vary as will the clock frequencies that use them. Such an oscillation signal would be detectable with atomic clocks for a large range of DM masses ($m  \lesssim 10^{-13}$ eV) and interaction strengths. Clock DM searches are naturally broadband, with mass range depending on the total measurement time and specifics of the clock operation protocols (see \cite{Tsai:2021lly} for details). A multidisciplinary team is needed to continue to develop atomic clocks as a tool for dark matter searches.

\subsection{Magnets}

Magnets are at the heart of all axion experiments and can be used in some searches for scalar-type dark matter. Advances in superconducting materials for the construction of magnets are enabling larger, higher-field magnets. The development of  production and quality of rare-earth barium copper oxide (REBCO) tapes is particularly promising and harnesses key technology that overlaps with other industries including fusion power. 

The DMNI experiments are based on conventional large magnet technology based on Nb$_{3}$Sn and/or NbTi technology. To move to higher masses, 30–50 $\mu$eV range, it is proposed to design a high-temperature superconducting (HTS) insert to create a maximum central magnetic field strength of 32 T over a 15 cm diameter. The National High Magnetic Field Laboratory (MagLab) has been developing higher field REBCO inserts with current designs reaching a maximum field of 45 T~\cite{Bai2020} and future designs aiming for even higher field strengths across smaller bore diameters.
To move to smaller axion masses, proposals like DMRadio-GUT~\cite{Brouwer:2022bwo} require more modest fields but large volumes and large bores. More complicated geometries may be advantageous in some measurements. Advances in HTS technology and the corresponding cryogenics are critical for the entire field. 

R\&D efforts in this area will allow for the development of cost-effective large-volume high-field magnet designs that will benefit many axion detection experiments and beyond. We seek to establish a framework by which experiments can create optimized magnetic field profiles based on the individual experimental needs, minimizing signal losses/leakage while still maximizing the science potential of axion searches in the highest possible magnetic field. We are interested in partnering with experts at national labs and industry in order to design, construct and then successfully implement these next generation of magnets for use in the search for wave-like dark matter.

\subsection{Resonant Systems}

The detection of wave-like dark matter often relies on resonant readout to amplify the wave-like signal. Cavities with sophisticated geometry and tuning mechanisms have long been used to allow scanning across frequencies corresponding to axion masses greater than 1\,$\micro$-eV. The move to higher frequency and masses results in smaller cavity volumes  and lower quality factors for ordinary metals. 
The move to higher frequency requires cavities to leverage new ideas from both accelerators and quantum information sciences~\cite{Kuo_2021}. 

For axion searches below 1\,$\micro$-eV, the transition to a lumped element detection model requires the development of high quality factor resonant circuits that allow tuning at cryogenic temperatures. The fundamental limits on the quality factors of such circuits is a subject of active research.

Resonant systems are not unique to axion searches. In particular, dark photon searches rely on similar resonant circuit and cavity readouts. This is fundamental R\&D which is important for wave dark matter searches but will find application in related fields including accelerator development.

\subsection{Cross-Disciplinary Collaborations}

A strong R\&D program grows from harnessing technological advancements for application in particle detectors. This process needs expertise from a broad range of disciplines. The technologies that are needed for wave dark matter searches range from novel materials to fabrication of complex devices. A particular need in the search for wave-like dark matter are interdisciplinary collaborations with experts in quantum measurement and control. Nurturing broad cross-disciplinary collaborations is key to a strong program searching for wave dark matter and high energy physics more broadly.

\section{Strong Theory Program} \label{sec:theory}

\subsection{Direct Impact on Experiments}
A strong collaboration between theory and experiment is one of the hallmarks of the wave-like dark matter community. It's also at the heart of the current and rapid development of the area.
This interaction goes far deeper than theorists proposing models that are than tested in experiments or experimentalists providing data that is then used to constrain models. 
Indeed theorists often conceptualize the experiments~\footnote{Of course there are also very nice examples of this in the wider dark matter community, e.g.~\cite{Goodman:1984dc} for WIMP detection. For a wider appreciation of the fruitful interaction between theory and experiment see also the TF09 white paper~\cite{Essig:2022yzw}.} and then often closely collaborate in the following stages.

An early, perhaps the earliest example of the strength of this interaction is the seminal paper by Pierre Sikivie~\cite{Sikivie:1983ip} that outlined the two main approaches to axion searches that are still pursued today, the axion helioscope and the axion haloscope, and which followed quickly upon the realization that axions are a good dark matter candidate~\cite{Preskill:1982cy,Abbott:1982af,Dine:1982ah}. This resulted amongst other things in today's ADMX collaboration with which Pierre Sikivie is still connected, cf, e.g.~\cite{ADMX:2021nhd}.

More recent examples of this fruitful collaboration are experiments such as CASPEr, DMRadio, MADMAX, IAXO and ALPHA that arose from theoretical concepts and direct collaboration with experiments~\cite{Sikivie:1983ip,Horns:2012jf,Jaeckel:2013eha,Graham:2013gfa,Budker:2013hfa,Chaudhuri:2014dla,Caldwell:2016dcw,Silva-Feaver:2016qhh,Irastorza:2011gs,Vogel:2013bta,Lawson:2019brd}. Similarly strong interactions exist not only for axions, but also for scalar and vector DM, cf.~\cite{Stadnik:2014ala,Graham:2015ifn,Berengut:2017zuo} for some examples and~\cite{Antypas:2022asj} for many more.

\subsection{Astrophysics, Cosmology and Phenomenology}
An important aspect of theoretical support for experiments is also to identify areas that are already excluded by other means as well as, more positively, detect areas that are especially promising because other experiments or observations have found hints that point into the respective area. 

This requires development of meaningful ``benchmark scenarios'' as well as the interpretation of a wide range of data from experiments and observation against these benchmarks.

A further, particularly important aspect for the very weakly coupled particles that are the subject of the wave-like dark matter community are astrophysical and cosmological observations. Due to their weak couplings stars in a broad range of evolutionary stages are outstanding probes of these particles and provide the best constraints (see the famous book~\cite{Raffelt:1996wa} as well as the discussion in~\cite{Adams:2022pbo} where more references can be found) as well as intriguing hints (cf., e.g.,~\cite{Giannotti:2015kwo,Giannotti:2017hny}). Also the propagation of photons through the cosmic medium, featuring magnetic fields, but also a plasma of charged particles as well as various types of photons (cosmic microwave photons, extra galactic background light etc.) is modified in the presence of axions and similar particles and therefore can provide powerful constraints but also intriguing hints, e.g.~\cite{DeAngelis:2007dqd,Simet:2007sa,Mirizzi:2009iz,Mirizzi:2009nq,Horns:2012kw,Meyer:2013pny,HESS:2013udx,Fermi-LAT:2016nkz,Zhang:2018wpc,Guo:2020kiq,Buehler:2020qsn,Cheng:2020bhr} (see also~\cite{Adams:2022pbo} for more details and references). Further progress will, of course, arise from the complementarity with new data from existing as well as future cosmic ray observatories such as HESS~\cite{HESS:2013udx}, Fermi-LAT~\cite{Fermi-LAT:2016nkz}, CTA~\cite{CTA:2020hii}, LHAASO~\cite{Long:2021udi} but also from new telescopes such as the newly active James Webb telescope~\cite{Gardner:2006ky} or the Vera Rubin Observatory~\cite{Mao:2022fyx}, see, again,~\cite{Adams:2022pbo} for an in depth discussion and references.

Cosmology, too, can provide powerful constraints. For example, the first papers on the misalignment can be interpreted as a cosmological constraint on the parameter space as certain parameter regions may produce too much dark matter inconsistent with observations~\cite{Preskill:1982cy,Abbott:1982af,Dine:1982ah}. Famously, the lower boundary on the mass of wave-like dark matter candidates $\sim 10^{-22}\,{\rm eV}$ also arises from their impact on cosmological structure formation~\cite{Hlozek:2014lca,Marsh:2015xka}. In particular for low masses CMB polarization measurements can also be a tool to probe axion and ALP dark matter, see~\cite{Harari:1992ea} as well as~\cite{Fedderke:2019ajk} for a more recent study.

An improved theoretical understanding of the cosmological production of dark matter may also give crucial information on where as well as how to search for dark matter. For example in a scenario where the last Peccei-Quinn symmetry breaking happens after inflation (so-called post inflationary scenario), the dark matter density is a definite function of the axion mass. Comparing with the rather precisely observed (average) density would allow to get a precise prediction for the axion mass. However, this determination is currently precluded by the insufficient understanding of the production that receives possibly significant  contributions from topological defects~\cite{Harari:1987ht,Davis:1989nj,Dabholkar:1989ju,Hagmann:1990mj,Battye:1993jv,Yamaguchi:1998gx,Chang:1998tb,Hagmann:2000ja,Sikivie:2006ni,Hiramatsu:2010yn,Hiramatsu:2010yu,Hiramatsu:2012gg,Hiramatsu:2012sc,Kawasaki:2014sqa,Gorghetto:2018myk,Martins:2018dqg,Hindmarsh:2019csc,Gorghetto:2020qws,Dine:2020pds,Hindmarsh:2021vih,Buschmann:2021sdq} in addition to the misalignment production.
The same scenario also suggests that dark matter may be inhomogeneously distributed on very small (e.g. solar system sized) scales, e.g. featuring miniclusters~\cite{Kolb:1993zz,Kolb:1993hw}. This has direct impact on the experiments (searching in resonant vs broadband mode)~\cite{Vaquero:2018tib} but also opens the potential for synergies with astrophysical observation such as gravitational lensing~\cite{Kolb:1995bu,Zurek:2006sy,Fairbairn:2017dmf,Fairbairn:2017sil} (see however also~\cite{Katz:2018zrn}). To bring this to fruition improvements in the quantitative understanding of the formation (cf., e.g.,~\cite{Kolb:1994fi,Kolb:1995bu,Zurek:2006sy,Fairbairn:2017sil,Enander:2017ogx,Vaquero:2018tib,Eggemeier:2019khm,Ellis:2020gtq,Ellis:2022grh}) as well as the survival of these structures~\cite{Tinyakov:2015cgg,Dokuchaev:2017psd,Kavanagh:2020gcy,Dandoy:2022prp} would be welcome.

\subsection{New Theoretical Target Areas}
Last but definitely not least, fundamental theory research can first of all be the original motivation for new (wave-like or other) dark matter candidates. It can open entirely new or previously overlooked areas of parameter space. It provides a framework to understand and interpret experimental results and makes and highlights meaningful connections to other areas of particle physics. Let us look at a few examples of this role in the following.

The first example is clearly the QCD axion itself~\cite{Peccei:1977hh,Peccei:1977ur,Weinberg:1977ma,Wilczek:1977pj}. Motivated by a theoretical finetuning problem of the SM (strong CP problem) it led to concrete predictions that were soon tested and excluded. This then gave rise to the advent of ``invisible'' axion models~\cite{Kim:1979if,Shifman:1979if,Zhitnitsky:1980tq,Dine:1981rt}.
As already mentioned, measurement of an axion electron coupling would point to a DFSZ model and therefore an extended Higgs sector that can be tested at the LHC or a future collider, thereby making a connection that would not be possible without a more complete theory model.

A recent example is the widening of parameter space for QCD axion models. Whereas the original models predicted a strict relation between the axion mass and the coupling to gluons (and also an order of magnitude relation to that with photons) recent theory  efforts~\cite{Hook:2014cda,Fukuda:2015ana,Chiang:2016eav,Dimopoulos:2016lvn,Gherghetta:2016fhp,Kobakhidze:2016rwh,Agrawal:2017ksf,Agrawal:2017evu,Gaillard:2018xgk,Hook:2018jle,Csaki:2019vte,Gupta:2020vxb,Gherghetta:2020ofz,DiLuzio:2021pxd,DiLuzio:2021gos} (see also~\cite{Rubakov:1997vp,Berezhiani:2000gh,Gianfagna:2004je,Hsu:2004mf} for  some older works) have led to the realization that a wider range of possibilities exist, albeit at the price of some model complications. Similarly the relation to the photon coupling was significantly widened~\cite{Kim:1998va, DiLuzio:2016sbl, DiLuzio:2017pfr, DiLuzio:2020wdo,Plakkot:2021xyx,Sokolov:2021ydn}. These motivates new experimental searches with a broad range of techniques.

On the more theoretical side, string theory provides significant motivation for light (pseudo-)scalars~\cite{Conlon:2006tq,Conlon:2006ur,Arvanitaki:2009fg,Cicoli:2012sz,Marsh:2015xka,Acharya:2015zfk,Visinelli:2018utg,Broeckel:2021dpz,Demirtas:2021gsq,Mehta:2020kwu,Mehta:2021pwf,Cicoli:2022lsq} as well as vectors~\cite{Dienes:1996zr,Lukas:1999nh,Abel:2003ue,Blumenhagen:2005ga,Abel:2006qt,Abel:2008ai,Goodsell:2009pi,Goodsell:2009xc,Goodsell:2010ie,Heckman:2010fh,Bullimore:2010aj,Cicoli:2011yh,Goodsell:2011wn}. The former often arise as moduli that feature a scalar and a pseudoscalar ``axionic'' component. 

String theory models also provides a strong connection to cosmology and the corresponding observations. An example is the prediction of a significant amount of axionic dark radiation arising from moduli decays~\cite{Cicoli:2012aq,Higaki:2012ar,Hebecker:2014gka,Angus:2014bia,Cicoli:2015bpq}. This provides constraints but can be consistent with observations~\cite{Cicoli:2022lsq} but is also a potential observable in cosmological observations as well as experiments such as IAXO~\cite{IAXO:2019mpb}.
Moreover, string theory also opens the opportunity to build more complete models that also include to inflation, again providing a connection to new observables~\cite{Cicoli:2022lsq} such as dark radiation constraints from future CMB experiments, e.g. CMB-S4~\cite{CMB-S4:2016ple}.

\section{Conclusions} \label{sec:conclusions}

While wavelike dark matter candidates (such as the QCD axion) have always been well-motivated, the last decade has seen explosive growth in both enthusiasm for WLDM models and search techniques, and fortuitous alignment between the enabling technologies for those searches and the strategic needs of the US quantum workforce.  In the coming decade, the WLDM community endorses supporting a strong program of experiments at different scales and R\&D in key enabling technologies.  The goal of this program is to enable a discovery of wavelike dark matter which will revolutionize our understanding of the universe and our place in it.


\section*{Acknowledgements}
JJ is grateful for support from the EU ITN HIDDEN (No 860881). The conveners thank the Department of Energy and National Science Foundation for their support of the US axion dark matter community.

\bibliographystyle{utphys}
\bibliography{CF2_Ref}

\end{document}